\documentclass[sigconf]{acmart}
\AtBeginDocument{%
  }

\copyrightyear{2026}
\acmYear{2026}
\setcopyright{cc}
\setcctype{by}
\acmConference[CIKM '26]{Proceedings of the 35th ACM International Conference on Information and Knowledge Management}{November 07--11, 2026}{Rome, Italy}
\acmBooktitle{Proceedings of the 35th ACM International Conference on Information and Knowledge Management (CIKM '26), November 07--11, 2026, Rome, Italy}
\acmDOI{10.1145/3799682.3840109}
\acmISBN{979-8-4007-2539-5/2026/11}

\usepackage{subcaption}
\usepackage{multirow}
\usepackage{array}
\usepackage{balance}
\usepackage{multicol}
\usepackage{makecell}
\usepackage{stfloats}
\usepackage{enumitem}
\setlist{itemsep=2pt, parsep=0pt, topsep=2pt}
\begin{document}

%%
%% The "title" command has an optional parameter,
%% allowing the author to define a "short title" to be used in page headers.
\title{SORT: A Systematically Optimized Ranking Transformer for Industrial-scale Recommenders}

%%
%% The "author" command and its associated commands are used to define
%% the authors and their affiliations.
%% Of note is the shared affiliation of the first two authors, and the
%% "authornote" and "authornotemark" commands
%% used to denote shared contribution to the research.
\author{Chunqi Wang}
\authornote{Corresponding author.}
\affiliation{
  \institution{Alibaba International Digital Commercial Group}
  \city{Hangzhou}
  \country{China}
}
\email{shiyuan.wcq@alibaba-inc.com}

\author{Bingchao Wu}
\affiliation{
  \institution{Alibaba International Digital Commercial Group}
  \city{Hangzhou}
  \country{China}
}
\email{wubingchao.wbc@alibaba-inc.com}

\author{Taotian Pang}
\affiliation{
  \institution{Alibaba International Digital Commercial Group}
  \city{Hangzhou}
  \country{China}
}
\email{pangtaotian.ptt@alibaba-inc.com}

\author{Jiahao Wang}
\affiliation{
  \institution{Alibaba International Digital Commercial Group}
  \city{Beijing}
  \country{China}
}
\email{wjh177423@alibaba-inc.com}

\author{Jie Yang}
\affiliation{
  \institution{Alibaba International Digital Commercial Group}
  \city{Beijing}
  \country{China}
}
\email{jiehong.yj@alibaba-inc.com}

\author{Jia Liu}
\affiliation{
  \institution{Alibaba International Digital Commercial Group}
  \city{Beijing}
  \country{China}
}
\email{lj410672@alibaba-inc.com}

\author{Hao Zhang}
\affiliation{
  \institution{Alibaba International Digital Commercial Group}
  \city{Beijing}
  \country{China}
}
\email{zh138764@alibaba-inc.com}

\author{Hai Zhu}
\affiliation{
  \institution{Alibaba International Digital Commercial Group}
  \city{Hangzhou}
  \country{China}
}
\email{zhuhai@alibaba-inc.com}

\author{Lei Shen}
\affiliation{
  \institution{Alibaba International Digital Commercial Group}
  \city{Hangzhou}
  \country{China}
}
\email{kenny.sl@alibaba-inc.com}

\author{Shizhun Wang}
\affiliation{
  \institution{Alibaba International Digital Commercial Group}
  \city{Beijing}
  \country{China}
}
\email{shaoan.wsz@alibaba-inc.com}

\author{Bing Wang}
\authornotemark[1]
\affiliation{
  \institution{Alibaba International Digital Commercial Group}
  \city{Hangzhou}
  \country{China}
}
\email{lingfeng.wb@alibaba-inc.com}

\author{Xiaoyi Zeng}
\affiliation{
  \institution{Alibaba International Digital Commercial Group}
  \city{Hangzhou}
  \country{China}
}
\email{yuanhan@alibaba-inc.com}

%%
%% By default, the full list of authors will be used in the page
%% headers. Often, this list is too long, and will overlap
%% other information printed in the page headers. This command allows
%% the author to define a more concise list
%% of authors' names for this purpose.
\renewcommand{\shortauthors}{Chunqi Wang et al.}

%%
%% The abstract is a short summary of the work to be presented in the
%% article.
\begin{abstract}
While Transformers have achieved remarkable success in LLMs through superior scalability, their application in industrial-scale ranking models remains nascent, hindered by the challenges of high feature sparsity and low label density. In this paper, we propose SORT (\textbf{S}ystematically \textbf{O}ptimized \textbf{R}anking \textbf{T}ransformer), a scalable model designed to bridge the gap between Transformers and industrial-scale ranking models. We address the high feature sparsity and low label density challenges through a series of optimizations, including request-centric sample organization, local attention, query pruning and generative pre-training. 
Furthermore, we introduce a suite of refinements to the tokenization, multi-head attention (MHA), and feed-forward network (FFN) modules, which collectively stabilize the training process and enlarge the model capacity.
To maximize hardware efficiency, we optimize our training system to elevate the model FLOPs utilization (MFU) to 45\%.
Extensive experiments demonstrate that SORT outperforms strong baselines and exhibits excellent scalability across data size, model size and sequence length, while remaining flexible at integrating diverse features. 
Finally, online A/B testing in large-scale e-commerce scenarios confirms that SORT achieves significant gains in key business metrics—including orders (+7.47\%), buyers (+6.67\%) and GMV (+8.65\%)—while simultaneously cutting latency by 62\% and boosting throughput nearly sevenfold (+589\%). SORT has been fully deployed in production, serving all users on AliExpress.

\end{abstract}

%%
%% The code below is generated by the tool at http://dl.acm.org/ccs.cfm.
%% Please copy and paste the code instead of the example below.
%%

%%
%% Keywords. The author(s) should pick words that accurately describe
%% the work being presented. Separate the keywords with commas.
%% A "teaser" image appears between the author and affiliation
%% information and the body of the document, and typically spans the
%% page.

% \received{10 February 2025}
% \received[revised]{1 May 2025}
% \received[accepted]{16 May 2025}

%%
%% This command processes the author and affiliation and title
%% information and builds the first part of the formatted document.
\settopmatter{printacmref=true}
\maketitle

% \newcommand\kddavailabilityurl{https://doi.org/10.5281/zenodo.15567745}

% \ifdefempty{\kddavailabilityurl}{}{
% \begingroup\small\noindent\raggedright\textbf{KDD Availability Link:}\\
% % please change the following context to include multiple artifacts if necessary.
% The source code of this paper has been made publicly available at \url{\kddavailabilityurl}.
% \endgroup
% }

\section{Introduction}

The Transformer architecture\cite{vaswani2017attention} has established dominance in large language models (LLMs)\cite{touvron2023llama,liu2024deepseek,yang2025qwen3}, owing to its excellent scalability. However, its application to ranking models in industrial-scale recommenders remains in a nascent stage. For a long time, the landscape of ranking models has been dominated by specialized networks that focus on explicit feature interaction. For instance, DIN\cite{zhou2018deep} is designed to extract patterns from sequential features, while DeepFM\cite{guo2017deepfm} and DCN\cite{wang2017deep} specialize in heterogeneous feature interaction.
These networks are primarily designed to capture feature interaction by low-compute operations including vector-wise operations and low-dimensional matrix multiplication, which struggle to leverage the massive parallel computational capacity of modern accelerators and are also unable to capture high-dynamic data patterns.
Given the critical role of ranking models in industrial recommenders and the huge success of LLMs, we propose rebuilding ranking models based on the Transformer architecture, paving the way for highly unified and scalable ranking models.

However, directly applying the Transformer architecture to ranking models is sub-optimal due to the inherent differences between recommendation and language modeling. A primary challenge lies in the synergy between \textbf{high feature sparsity} and \textbf{low label density}, which hinders the model's ability to learn robust representations from discrete, massive feature spaces with infrequent supervisory signals.
In language modeling, feature sparsity is relatively low due to the adoption of subword tokenization techniques such as BPE\cite{sennrich2016neural} and wordpiece\cite{song2021fast}, which constrain the vocabulary size to a manageable range (e.g., within hundreds of thousands). In the meanwhile, its label density is inherently high, as every position within the text sequence possesses a supervisory signal (i.e., next-token prediction) drawn from the entire vocabulary-sized search space. 
Conversely, ranking models must handle item vocabularies at the billion scale, necessitating enormous embedding tables and resulting in extremely high feature sparsity. 
% This challenge is compounded by low label density: while the user historical sequence occupies the vast majority of the input context, it lacks explicit labels at these positions. Unlike the dense feedback in training LLMs, training ranking models typically relies on sparse binary signals (e.g., click or purchase) associated exclusively with the exposed target items.
The challenge is further defined by low label density, stemming from two key factors. First, the historical sequence makes up the majority of the input but lacks direct supervision, which is restricted solely to the target items. Second, these signals are binary and sparse by nature, where the vast majority of samples are negative (zero-valued), such as instances of non-clicks.
This information asymmetry, where a massive parameter space is regularized by sparse binary labels, not only induces overfitting, but also leads to substantial computational waste on long historical sequences.

To address these issues, we introduce \textit{request-centric sample organization}, \textit{local attention}, and \textit{query pruning} to mitigate the computational inefficiency caused by the long historical sequences. By reallocating the Transformer's computation to label-relevant interactions, these techniques significantly reduce computational overhead. In addition, we incorporate \textit{generative pre-training} to resolve the issues introduced by the information asymmetry of high feature sparsity and low label density. Since the sparse binary signals from target items are insufficient to regularize a billion-scale parameter space, this approach augments label density through generative training (i.e., next-item prediction), thereby effectively alleviating overfitting and stabilizing the training of Transformer-based ranking models.
Complementing these optimizations, we further refine the Transformer architecture by enhancing the tokenization, multi-head attention layer (MHA) and feed-forward network (FFN) modules. For the tokenization module, we introduce \textit{special tokens} as feature boundaries. Within the MHA layer, we introduce \textit{QKNorm} and \textit{attention gate} to stabilize the training process. For the FFN module, we replace the standard architecture with a mixture-of-experts (MoE) layer, thereby increasing model capacity without incurring additional computational costs.

We refer to the proposed approach as SORT, standing for\break \textbf{S}ystematically \textbf{O}ptimized \textbf{R}anking \textbf{T}ransformer. Extensive experiments show that SORT achieves significant performance gains compared with strong baselines. We also evaluate SORT's scalability across multiple aspects: data size, model size and sequence length, demonstrating that SORT exhibits excellent scalability.
% Feature engineering is another critical aspect of ranking models. We demonstrate that SORT is flexible to integrate diverse handcrafted features, including multi-modal features, statistical features and other model-derived features.
Finally, extensive online A/B testing across multiple scenarios on AliExpress demonstrates that SORT not only yields substantial business growth (+7.47\% in orders, +6.67\% in buyers, and +8.65\% in GMV) but also drastically enhances serving efficiency, cutting latency by 62\% and boosting throughput nearly sevenfold (+589\%).

\section{Related Work}
Conventional deep learning recommendation models (DLRMs) for recommendation typically rely on specialized architectures to handle sequence features (e.g., DIN\cite{zhou2018deep}) and extensive handcrafted heterogeneous features (e.g., DeepFM\cite{guo2017deepfm}, DCN\cite{wang2017deep}). Unlike Transformers, these architectures lack scalability, resulting in limited model capacity. Consequently, patterns are primarily captured through the rote memorization of sparse embeddings (e.g., item embeddings). Previous studies have demonstrated that such massive embedding tables often impair generalization and lead to overfitting\cite{zhang2022towards,wang2025scaling}.

Motivated by LLMs, HSTU\cite{zhai2024actions} introduces scalable architectures into RMs. Specifically, HSTU treats ranking as a next-token prediction task and relies solely on a unified sequence as input. This modification amortizes the training cost since it allows processing multiple candidates within a single sample. Subsequent models, such as MTGR\cite{han2025mtgr}, GenRank\cite{huang2025towards} and this work, have also adopted a similar paradigm.
Wukong\cite{zhang2024wukong} also introduces scalable architectures into ranking tasks. However, in contrast to HSTU, Wukong focuses on handling heterogeneous feature interaction. RankMixer\cite{zhu2025rankmixer} also falls into this category.
Recently, OneTrans\cite{zhang2025onetrans} introduced a Transformer variant that leverages independent parameters to model heterogeneous features and user sequence. The independent parameters can be regarded as rule-based routing MoE.
Different from that, SORT avoids rule-based routing MoE instead using a data-driven routing one.
To better position our work, one can envision a spectrum where traditional DLRMs—reliant on specialized feature engineering—occupy one end, and LLMs—characterized by unified, scalable architectures—occupy the other. Compared to prior efforts, SORT shifts significantly toward the LLM side of this spectrum. By embracing a more streamlined architecture, SORT departs from the complexity of domain-specific interaction modules, paving the way for a more generalized ranking paradigm.

Despite these architectural advancements, GPSD\cite{wang2025scaling} identified that the presence of sparse item ID features continues to hinder the full realization of model scalability. To address this, GPSD proposed a framework that employs generative pre-training and sparse embedding freezing strategy. This framework effectively unlocks the scalability of Transformers on ranking tasks while simultaneously mitigating the one-epoch overfitting problem\cite{zhang2022towards}. Therefore, we also incorporate GPSD in this work and prove its great effectiveness.

\section{SORT}
\begin{figure*}[htbp]
    \centering
    \includegraphics[width=0.8\textwidth]{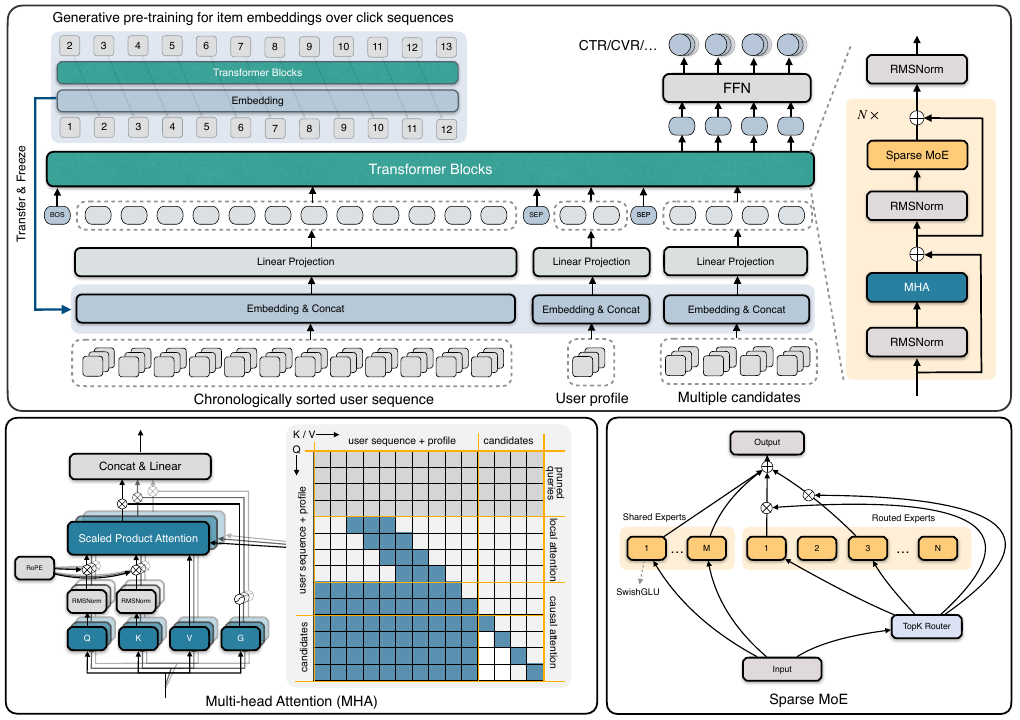}
    \caption{Overview of SORT. The input concatenates user history $\mathcal{H}$, profile $\mathcal{U}$, and candidates $\mathcal{C}$ with \texttt{BOS}/\texttt{SEP} tokens. Each block uses QKNorm and gated attention with a sparse mask, followed by a sparse MoE layer; query pruning progressively shortens the sequence across layers. The ranking head outputs multi-objective predictions on candidate tokens.}
    \label{fig:sort_arch}
\end{figure*}

\subsection{Preliminaries}
\subsubsection{Task}
We formulate the ranking problem as a multi-objective binary classification task to simultaneously predict the probabilities of three primary user behaviors: click, cart, and purchase. We use AUC to measure the model performance.

\subsubsection{Data}
\label{sec:data}
Conventional ranking models typically operate on impression-centric samples (i.e., one candidate per sample).
Under this setting, the model redundantly re-processes request-invariant features (e.g., user profiles and sequences) for each intra-request candidate, leading to unnecessary computational overhead.
To optimize efficiency, we implement a request-centric data organization that encapsulates multiple intra-request candidates into a unified structure.
Specifically, a request $r$ involves a candidate set $\mathcal{C} = \{c_j\}_{j=1}^N$ to be ranked, where each item $c_j$ is characterized by item metadata, statistical features, and user-item cross features.

Simultaneously, the request $r$ encompasses request-invariant features that are shared across all candidates in $\mathcal{C}$.
These features consist of two distinct components:
\begin{itemize}
    \item User sequence ($\mathcal{H}$): A chronological sequence of historical behaviors $\mathcal{H} = (i_1, i_2, \dots, i_L)$, where each behavior $i_t$ incorporates static item metadata and the associated interaction context (e.g., timestamp, action type, and scene).
    \item User profiles ($\mathcal{U}$): A combination of features integrating static user metadata and statistical features.
\end{itemize}
Instead of forming $N$ impression-centric samples $\{ \langle \mathcal{H}, \mathcal{U},  c_{j} \rangle \}_{j=1}^N$ for the request $r$, 
we define a unified request-centric sample $\mathcal{S}$ as follows:
$$\mathcal{S} = \langle \mathcal{H}, \mathcal{U}, \mathcal{C} \rangle$$

\subsection{Model}
Our proposed model is built upon the Transformer architecture, systematically optimized for industrial ranking. Figure \ref{fig:sort_arch} provides an overview of the model.

Departing from the first Transformer\cite{vaswani2017attention}, SORT directly integrates evolved best practices in LLMs\cite{touvron2023llama}: Pre-Normalization\cite{xiong2020layer} with RMSNorm \cite{zhang2019root}, RoPE\cite{su2024roformer} for relative positioning, and Swish-GLU\cite{shazeer2020glu} activations.
Other key optimizations are detailed below.

\subsubsection{Tokenization}
% special token
In the tokenization phase, we generate a single token for each item (including both items in the user's historical sequence and candidate items), while the user profile is mapped to multiple tokens. 
We also experimented with tokenizing candidate items into multiple tokens but a negligible performance difference was observed.

The tokenization pipeline consists of performing embedding lookup on multiple raw features, followed by concatenation, and finally applying linear projection and normalization to map the result to the target model dimension.

In addition to these feature-derived tokens, we introduce two special tokens: a special token, \texttt{BOS}, is prepended to the user history sequence, and two \texttt{SEP} tokens are inserted between distinct feature groups. 
These special tokens serve a dual purpose: they provide positional information to help the model identify sequence boundaries (i.e., the start and end), and simultaneously act as attention sinks\cite{xiao2023efficient}, absorbing excess attention scores.
We observe that adding these special tokens significantly enhances model performance.

The final input sequence $\mathbf{S}$ is constructed by concatenating all aforementioned tokens in a specific order: $$\mathbf{S} = [\texttt{BOS} ; \text{Tokenize}(\mathcal{H}) ; \texttt{SEP} ; \text{Tokenize}(\mathcal{U}) ; \texttt{SEP} ; \text{Tokenize}(\mathcal{C})]$$

\subsubsection{Multi-head Attention}
Multi-head Attention serves as the core operation of the Transformer, establishing direct dependencies between distinct tokens based on semantic similarity.
We incorporate RoPE\cite{su2024roformer} to explicitly model relative positional information.
To ensure candidates are processed independently, we apply a diagonal mask on the candidates and assign them an identical position ID, so that each candidate attends only to itself and the preceding history/profile tokens.
To enhance training stability and model performance, we implement QKNorm\cite{henry2020query}, which applies an RMSNorm\cite{zhang2019root} layer to the queries and keys. Furthermore, drawing inspiration from HSTU\cite{zhai2024actions} and gated attention\cite{qiu2025gated}, we introduce a gating operation following the scaled dot-product attention, a modification that yields further performance gains.

To mitigate the computational overhead associated with long user sequences, we utilize a sparse attention mask. Building upon the standard causal mask, we restrict the attention mechanism for the majority of tokens in the historical behavior sequence to local attention\cite{beltagy2020longformer} while preserving standard causal attention for all candidate tokens and extra few tokens near the candidates. This design is highly conducive to processing very long user sequences, reducing the $\mathcal{O}(L^2)$ time complexity to $\mathcal{O}(W\cdot L)$, where $L$ is the sequence length and $W$ is the window size.

Finally, we leverage a query pruning strategy similar to OneTrans \cite{zhang2025onetrans}. At each layer, we prune queries associated with tokens distant from the candidate tokens, resulting in progressively shorter sequences toward the top layers. At the final layer, we keep at most 128 non-candidate tokens. This modification almost halves the computational cost. Moreover, we see that this approach leads to unexpectedly better model performance compared to the non-pruning baseline, which can be attributed to the inductive bias of the inherent temporal decay in recommendation systems.

% Here we provide a formal description of the $l$-th layer. Let $L^{(l-1)}$ and $L^{(l)}$ denote the sequence lengths of the input and output for this layer, respectively, where $L^{(l)} \le L^{(l-1)}$ due to the layer-wise query pruning. For each head $i$ with dimension $d_k$, given the input $\mathbf{X} \in \mathbb{R}^{L^{(l-1)} \times d}$, we calculate:
% $$\begin{aligned}
% \mathbf{Q}_{i} &= \text{RMSNorm}(P(\mathbf{X},L^{(l)})\mathbf{W}^Q_{i}) \\
% \mathbf{K}_{i} &= \text{RMSNorm}(\mathbf{X}\mathbf{W}^K_{i}) \\
% \mathbf{V}_{i} &= \mathbf{X}\mathbf{W}^V_{i} \\
% \mathbf{G}_{i} &= \sigma(P(\mathbf{X},L^{(l)})\mathbf{W}^G_{i})
% \end{aligned}$$
% where $\mathbf{W}^Q_{i}, \mathbf{W}^K_{i}, \mathbf{W}^V_{i} \text{ and } \mathbf{W}^G_{i} \in \mathbb{R}^{d \times d_k}$ are trainable parameters, $\sigma$ denotes the sigmoid operator, and $P(\mathbf{X}, n)$ is a function that extracts the last $n$ rows of matrix $\mathbf{X}$. Then, the output of this head is calculated by
% $$\text{Head}_i = \mathbf{G}_{i} \odot \left( \text{Softmax}\left( \frac{\mathcal{R}(\mathbf{Q}_{i}, \mathbf{K}_{i}) + \mathbf{M}}{\sqrt{d_k}} \right) \mathbf{V}_{i}\right)$$
% where $\mathcal{R}(\cdot)$ denotes the RoPE relative position encoding, and $\mathbf{M} \in \{0, -\infty\}^{L^{(l)} \times L^{(l-1)}}$ is the attention mask. Finally, outputs of all $h$ attention heads are concatenated and linear projected by $W^O\in\mathbb{R}^{hd_k \times d}$ to form the final output of this layer:
% $$\text{Output} = [\text{Head}_1 ; \text{Head}_2 ; \dots ; \text{Head}_h] \mathbf{W}^O$$

\subsubsection{Feed-Forward Network}
The Feed-Forward Network (FFN) constitutes another critical component of the Transformer architecture, designed to enhance the network's non-linearity and capacity. 
Although the original Transformer\cite{vaswani2017attention} adopts ReLU activation function, later Transformers commonly use SwishGLU\cite{shazeer2020glu} for better performance, which is also adapted in our implementation.
Another prevailing strategy for optimizing the FFN module is the sparse Mixture-of-Experts (MoE) framework. This approach deploys multiple experts and employs a routing network to assign routing weights, thereby activating only a specific subset of top-k experts for each input. The final output is obtained via the weighted aggregation of the selected experts' outputs. The primary advantage of Sparse MoE lies in its ability to significantly scale up the number of parameters—thereby enhancing the model's memorization capacity—without incurring additional computational overhead. This characteristic is particularly critical in recommendation scenarios, which demand strict adherence to low-latency constraints.

We conducted a comparative analysis between Switch-style MoE \cite{fedus2022switch} and DeepSeek-style MoE\cite{dai2024deepseekmoe}. The latter not only obviates the need for tuning the auxiliary load balancing loss hyperparameter but also delivers slightly better performance as shown in Section~\ref{sec:moe_sparsity_ratio}. Consequently, we adopt DeepSeek-style MoE in our final architecture.

\subsubsection{Ranking Head and Loss Function}
% 我们的模型被设计为一个多目标预测模型。为了得到排序分数，我们简单地在候选项token对应的输出上使用一个MLP。MLP有一个ReLU函数激活的隐层，最终输出层使用sigmoid函数进行激活得到CTR和CVR等分数。每个目标我们使用常规的交叉熵损失函数，所有目标的损失加权求和得到最终的损失。
To derive the final ranking scores, we apply an FFN to the output hiddens corresponding to the candidate item tokens. This FFN comprises a hidden layer with ReLU activation, followed by an output layer utilizing a sigmoid activation function to yield prediction for each objective. We employ a binary cross-entropy loss for each objective. The total training loss is computed as the weighted summation of the losses from all individual objectives.

\subsubsection{Sparse Embedding Strategy}
Feature sparsity presents a critical challenge in recommendation models, primarily stemming from the massive scale of users and items. Incorporating learnable embedding tables for these entities introduces an enormous number of parameters, which frequently leads to severe model overfitting\cite{zhang2022towards}.

To address this issue, we adopt distinct strategies for user and item representation. For users, we eschew explicit user id feature and instead characterize users through their historical behavior sequences and profile (e.g., age, location). 
For items, we adopt a generative pre-training paradigm proposed in GPSD\cite{wang2025scaling} to mitigate sparse-embedding overfitting by augmenting label density via next-item supervision. We investigate two strategies: \textit{sparse transfer} and \textit{sparse transfer + sparse freeze}. The first involves initializing with pre-trained sparse embeddings and fine-tuning them during the training of the ranking model; the second additionally freezes the sparse embeddings throughout the training process.
% For items, we incorporate the generative pre-training paradigm proposed in GPSD\cite{wang2025scaling}. Specifically, we train a next-item prediction model based on user click sequences and transfer the learned sparse parameters, i.e. the item embedding table, into the ranking model. Crucially, this item embedding table remains frozen throughout the training of the ranking model. This strategy effectively mitigates overfitting and yields significant improvements in model performance.

\subsection{Infrastructure}
Our training system is built upon RecIS~\cite{zong2025recis}, with the MoE module implemented via Megablocks\cite{gale2023megablocks}. The architecture comprises a sparse module bottlenecked by I/O and memory access, and a dense module bottlenecked by computation. For the sparse module, beyond conventional optimizations (operator fusion, merged memory accesses, dynamic embedding), we introduce multi-process-group communication (MPGC), which schedules feature-embedding traffic across parallel process groups to hide communication latency. For the dense module, since existing attention operators including SDPA\cite{pytorch_sdpa} and FlashAttention\cite{dao2022flashattention} are ill-suited to our structurally complex sparse masks, we develop a general sparse attention operator that uses block-wise computation with mask preloading to skip fully masked blocks, complemented by mixed-precision computing and gradient accumulation. Together, these optimizations elevate MFU from 13\% to 22\% for models at the base scale. Notably, for the large model, MFU further reaches 45\%, demonstrating that our training system scales efficiently with model size.

For inference, we convert dynamic graphs to static graphs via \texttt{torch.export} and AOTInductor\cite{pytorch_aotinductor}, and apply three classes of enhancements: (i) high-performance sparse attention kernels tailored to diverse masks, (ii) operator fusion that consolidates self-attention linear layers and merges fragmented trivial operators, and (iii) general LLM-style optimizations including half-precision computing, KV caching, and multi-context/multi-stream execution. Together, these deliver a 29.4\% throughput increase and a 29.3\% latency reduction over the unoptimized baseline.

\section{Experiments}
\subsection{Settings}
\subsubsection{Dataset}
\label{sec:dataset}
We collect the training set from the user logs of an industrial recommender system over the past months and reserve the subsequent day for evaluation. 
The dataset statistics are listed in Table \ref{tab:dataset}.

\begin{table}
    \begin{tabular}{lccc}
        \toprule
         & \#Requests & \#Users & \#Impressions \\
        \midrule
        Train & 0.6B & 50M & 9B \\
        Test & 20M & 4M & 0.3B \\
        \bottomrule
    \end{tabular}
    \caption{Dataset statistics.}
    \label{tab:dataset}
\end{table}

\subsubsection{Baseline}
% 我们选取了几个强baseline模型，包括Transformer,HSTU和OneTrans。
% 其中Transformer网络我们参考了GPSD中使用的版本进行了少量改动，具体的架构图见附录中的图1所示。
We select several strong baseline models, including Transformer, HSTU\cite{zhai2024actions} and OneTrans\cite{zhang2025onetrans}. The baseline Transformer is a standard implementation that lacks the specialized features of SORT, including local attention, query pruning, MoE, attention gate, and QKNorm. We carefully set the size of each model for fair comparisons, as listed in Table \ref{tab:model_scale}.

\begin{table}
    \scalebox{0.85}{\begin{tabular}{llcccc}
        \toprule
        Scale & Model & Depth & Width & Intermediate Size \\
        \midrule
        \multirow{4}{*}{Base}&Transformer & 6 & 512 & 1280 \\
        &OneTrans & 6 & 512 & 1280 \\
        &HSTU & 6 & 768 & - \\
        &SORT & 6 & 512 & 1280 \\
        \midrule
        \multirow{2}{*}{Small}&Transformer & 4 & 256 & 640 \\
        &SORT & 4 & 256 & 640 \\
        \midrule
        \multirow{2}{*}{Large}&Transformer & 12 & 1024 & 2560 \\
        &SORT & 12 & 1024 & 2560 \\
        \bottomrule
    \end{tabular}}
    \caption{Model settings. Because SORT is built on MoE, the effective intermediate size is the product of the number of the activated experts for each token and the intermediate size of each expert. HSTU lacks an FFN block, so its intermediate size is omitted in the table. To match the parameter number and FLOPs of other models, we accordingly widen HSTU.}
    \label{tab:model_scale}
\end{table}

\subsubsection{Training Settings}
We use AdamW optimizer to train all the models. We set $(\beta_1, \beta_2)=(0.9, 0.99)$ and weight decay to 0.01. We use learning rates of 5e-4, 2e-4 and 1e-4 for small, base and large models respectively. We use a batch size of 12K in all experiments.
Unless otherwise specified, we train each model by one epoch with a max user historical sequence length of 1K.
Except for retaining FP32 precision for the ranking head and MoE routers, we use automatic BF16 mixed precision to train the models.
Each model is trained on 4 to 32 A100 GPUs for 1 to 10 days, depending on the model size.

\subsection{Effectiveness of Generative Pre-training}
\label{sec:gp-experiment}
\begin{figure}[t] % t 表示放在页面顶部
    \centering
    \includegraphics[width=0.8\linewidth]{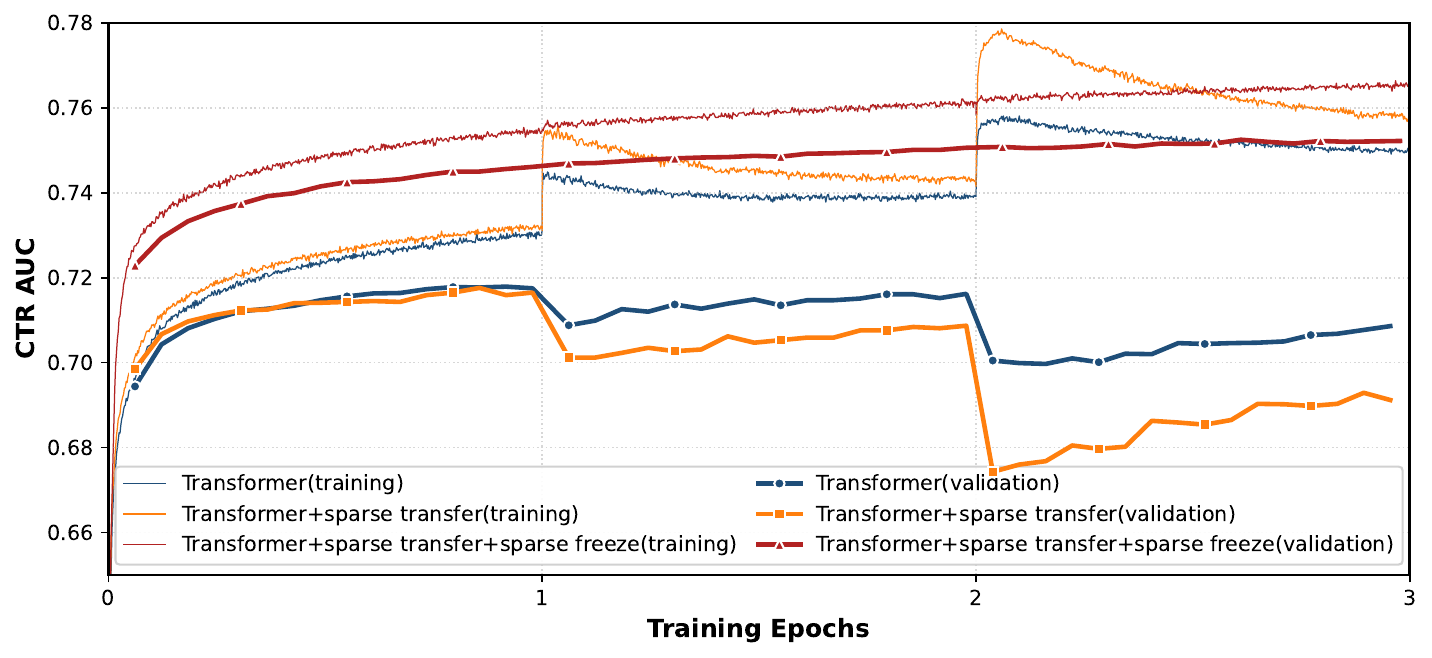}
    \caption{Training and validation AUC curves across multiple epochs under different strategies.}
    \label{fig:multi_epoch_training}
\end{figure}

We verified the effectiveness of the generative pre-training paradigm on the baseline Transformer architecture.
The results are presented in Table~\ref{tab:gp}. It can be observed that the strategy of simply transferring pre-trained embeddings (\textit{sparse transfer}) yields no performance improvements but even led to a slight degradation. In contrast, additionally freezing the pre-trained embeddings (\textit{sparse freeze}) yields a significant performance boost.
Another advantage of \textit{sparse freeze} is that it allows multi-epoch training without severe overfitting. As we can see for the CTR task in Figure~\ref{fig:multi_epoch_training}, the training and validation AUC curves both keep growing smoothly across multiple epochs.
Given the effectiveness, we adopt \textit{sparse transfer + sparse freeze} as the default strategy for the remainder of this paper.

\begin{table}
    \scalebox{0.75}{
    \begin{tabular}{lcccccc}
        \toprule
        \multirow{2}{*}{Model}&\multicolumn{2}{c}{CTR}&\multicolumn{2}{c}{CVR}&\multicolumn{2}{c}{AddCart} \\ 
        \cline{2-7}
        &AUC&Imp&AUC&Imp&AUC&Imp \\ 
        \midrule
        \makecell[l]{Transformer\\(from scratch)}&0.7175&-&0.8959&-&0.8455&- \\
        \hline
        \makecell[l]{Transformer\\+sparse transfer}&0.7162&-0.13pt&0.8943&-0.16pt&0.8449&-0.06pt \\
        \hline
        \makecell[l]{Transformer\\+sparse transfer\\+sparse freeze}&\textbf{0.7456}&\textbf{+2.81pt}&\textbf{0.9093}&\textbf{+1.34pt}&\textbf{0.8663}&\textbf{+2.08pt} \\
        \bottomrule
    \end{tabular}
    }
    \caption{Impact of generative pre-training. Imp stands for absolute AUC improvements.}
    \label{tab:gp}
\end{table}

\subsection{Hyperparameter Exploration}
\begin{figure}[t] % t 表示放在页面顶部
    \centering
    % 第一张子图
    \begin{subfigure}{\linewidth}
        \centering
        \includegraphics[width=0.85\linewidth]{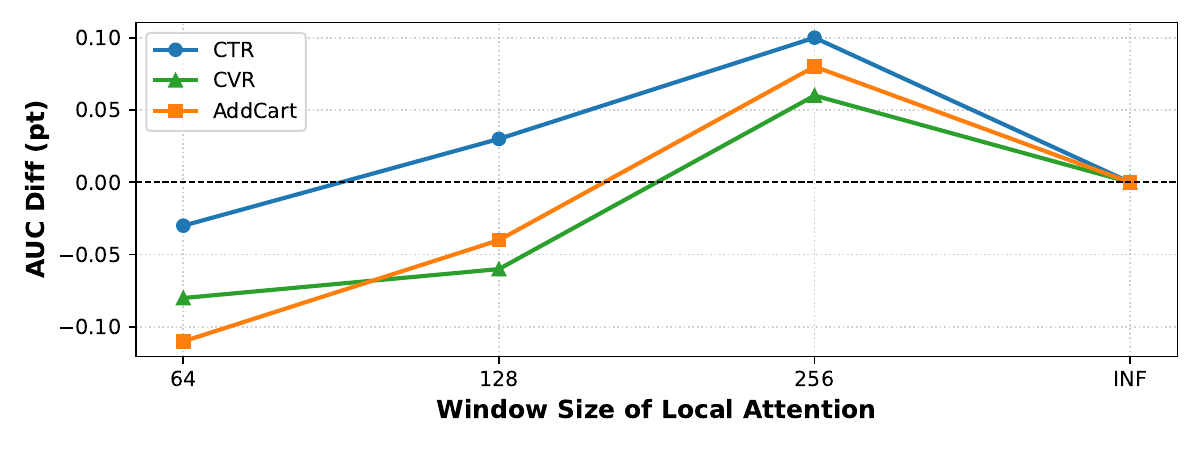}
        % \caption{Impact of local attention window size. A window size of \texttt{INF} (infinity) corresponds to the use of standard causal attention.}
        % \label{fig:attwindow_lift}
    \end{subfigure}
    \hfill % 在两张图之间插入弹性间距
    % 第二张子图
    \begin{subfigure}{\linewidth}
        \centering
        \includegraphics[width=0.85\linewidth]{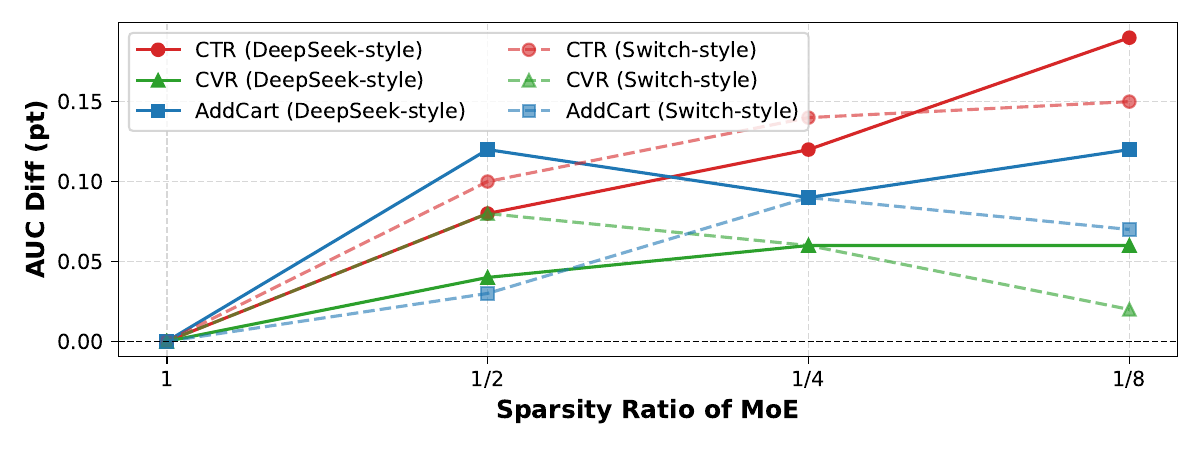}
        % \caption{Impact of MoE sparsity ratio. A sparsity ratio of 1 denotes that the model uses dense FFN.}
        % \label{fig:moe_lift}
    \end{subfigure}

    \caption{Impacts of various hyperparameter options.}
    \label{fig:combined_lift}
\end{figure}

In this section, we explore two important hyperparameters, i.e. window size of local attention and sparsity ratio of MoE, trying to figure out their optimal values. The results are presented in Figure~\ref{fig:combined_lift}.

\subsubsection{Local Attention Window Size} 
This hyperparameter controls how many tokens a token can at most attend to. If the value is small, it can only attend to very few tokens nearby thus lacks long-term dependencies. Conversely, if the value is large, it can attend to too many tokens thus cost much computation. 
We select three values for this hyperparameter: 64, 128, and 256.
It is evident that 256 achieves the superior results across all tested values. Surprisingly, its performance even surpasses that of the vanilla causal attention, which suggests that an appropriately constrained local window acts as a beneficial inductive bias, filtering out distant noise and allowing the model to focus on the most semantically relevant local context.

\subsubsection{MoE Sparsity Ratio}
\label{sec:moe_sparsity_ratio}
This hyperparameter is a ratio calculated by $$\frac{\text{The number of activated experts per token}}{\text{Total number of experts}},$$ which controls how sparse the MoE is. A dense model without MoE can be regarded as a special case where the sparsity ratio equals 1. We select three values for this hyperparameter: $1/2$, $1/4$, and $1/8$. Besides, we also compare two popular MoE implementations: Switch-style MoE and DeepSeek-style MoE.
In summary, the DeepSeek-style MoE demonstrates more effectiveness compared to the Switch-style counterpart. Notably, it simplifies the training process by obviating the effort required for hyperparameter tuning. The optimal result is observed at a sparsity ratio of 1/8.

\subsection{Architectural Ablation and Comparison}
\begin{figure}
    \centering
    \begin{subfigure}[b]{0.4\columnwidth}  % 使用 columnwidth 且值小于 0.5
        \centering
        \includegraphics[width=\textwidth]{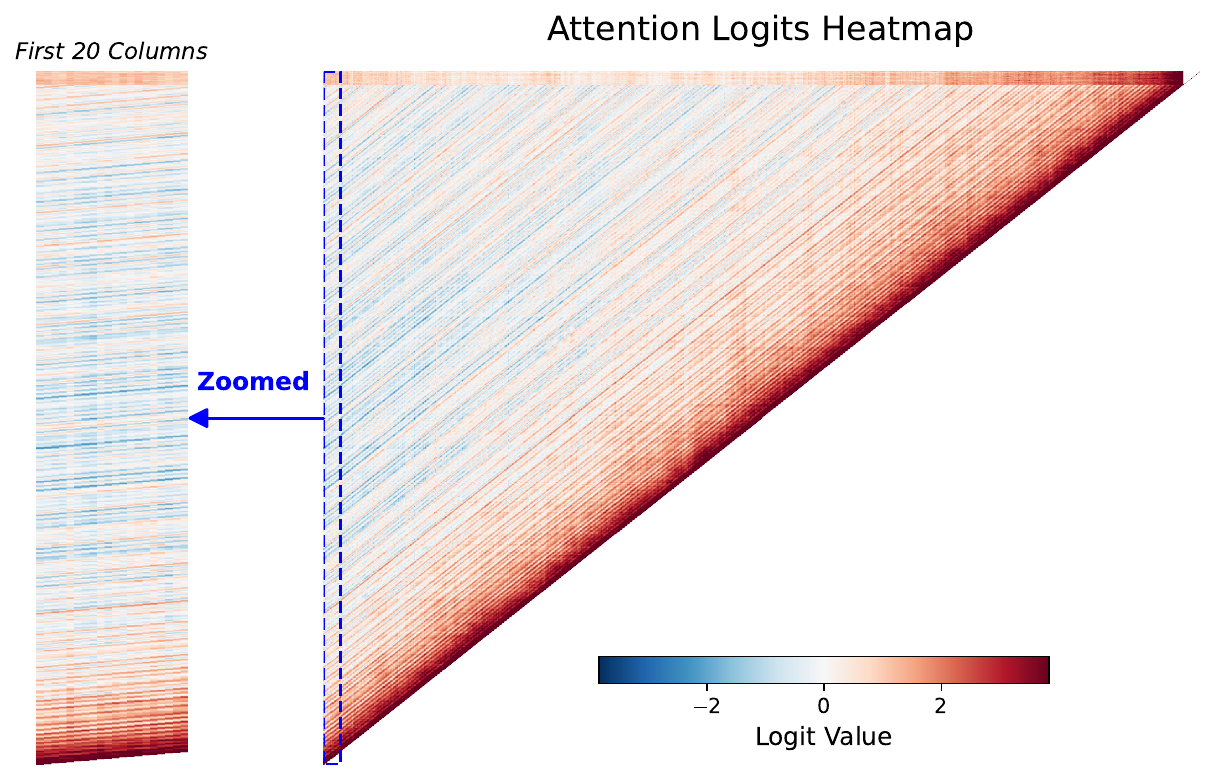}
        \caption{Transformer}
        \label{fig:main}
    \end{subfigure}
    % \hfill
    \begin{subfigure}[b]{0.4\columnwidth}
        \centering
        \includegraphics[width=\textwidth]{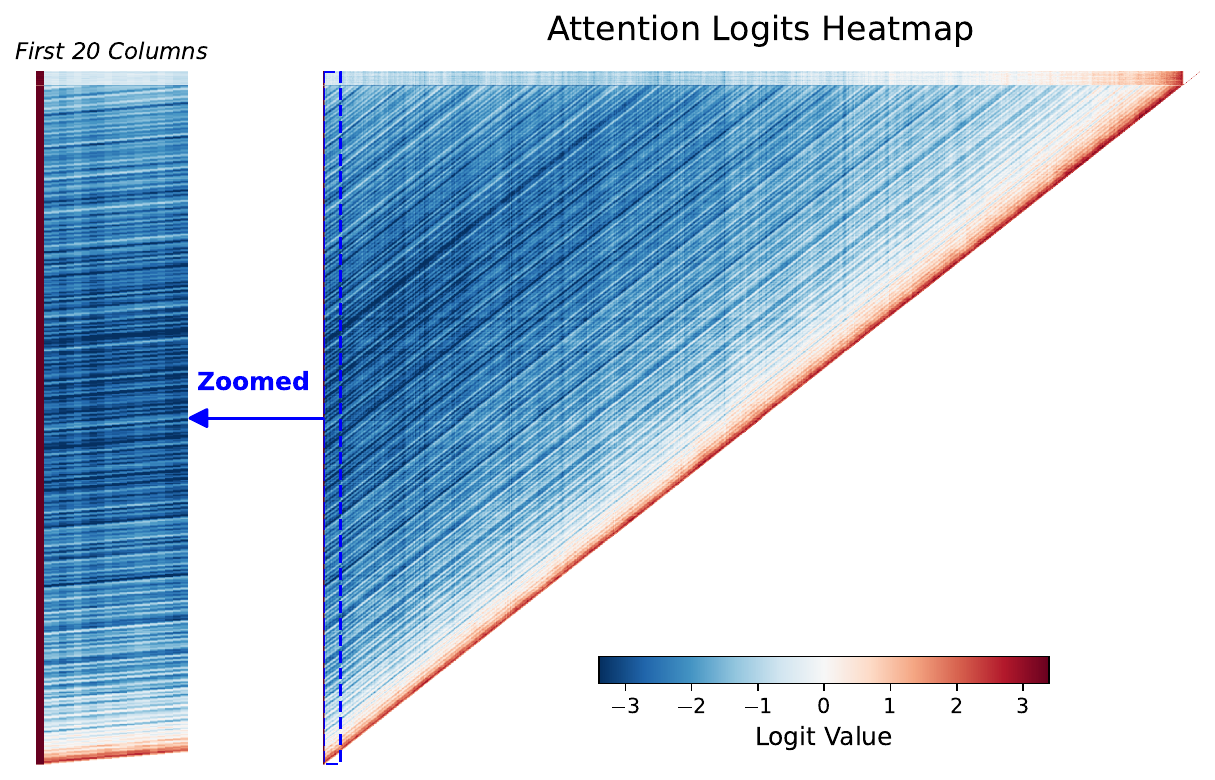}
        \caption{Transformer + ST}
        \label{fig:zoom}
    \end{subfigure}
     \caption{Heatmap visualization of the impact of special token on averaged attention logits at layer 0.}
    \label{fig:st_heatmap}
\end{figure}

We validate the efficacy of individual optimizations integrated into the base Transformer. As illustrated in Table~\ref{tab:main}, key techniques including \textit{special tokens}, \textit{query pruning}, and \textit{MoE}, demonstrate substantial independent gains, with \textit{special tokens} and \textit{query pruning} notably boosting CTR-AUC by +0.33pt and +0.26pt respectively in the base scale.
To visualize the impact of \textit{special tokens}, we show the attention heatmap in Figure \ref{fig:st_heatmap}, which demonstrates that the \texttt{BOS} token absorbs considerable attention weights, acting as attention sink\cite{xiao2023efficient}. We also observe that \textit{QKNorm} effectively mitigates the fluctuations of the attention weights, which is beneficial for the stability of the model training.
Other optimizations, such as \textit{attention gate} and \textit{local attention}, also demonstrate substantial gains.
The synergy of these optimizations culminates in our SORT model, which achieves state-of-the-art results across all primary tasks, significantly outperforming both standard Transformer and specialized architectures like HSTU and OneTrans. Notably, while \textit{local attention} only reduces FLOPs from 43G to 40G at a sequence length of 1K, it becomes crucial for longer sequences.

\begin{figure}[t] % t 表示放在页面顶部
    \centering
    \includegraphics[width=0.8\linewidth]{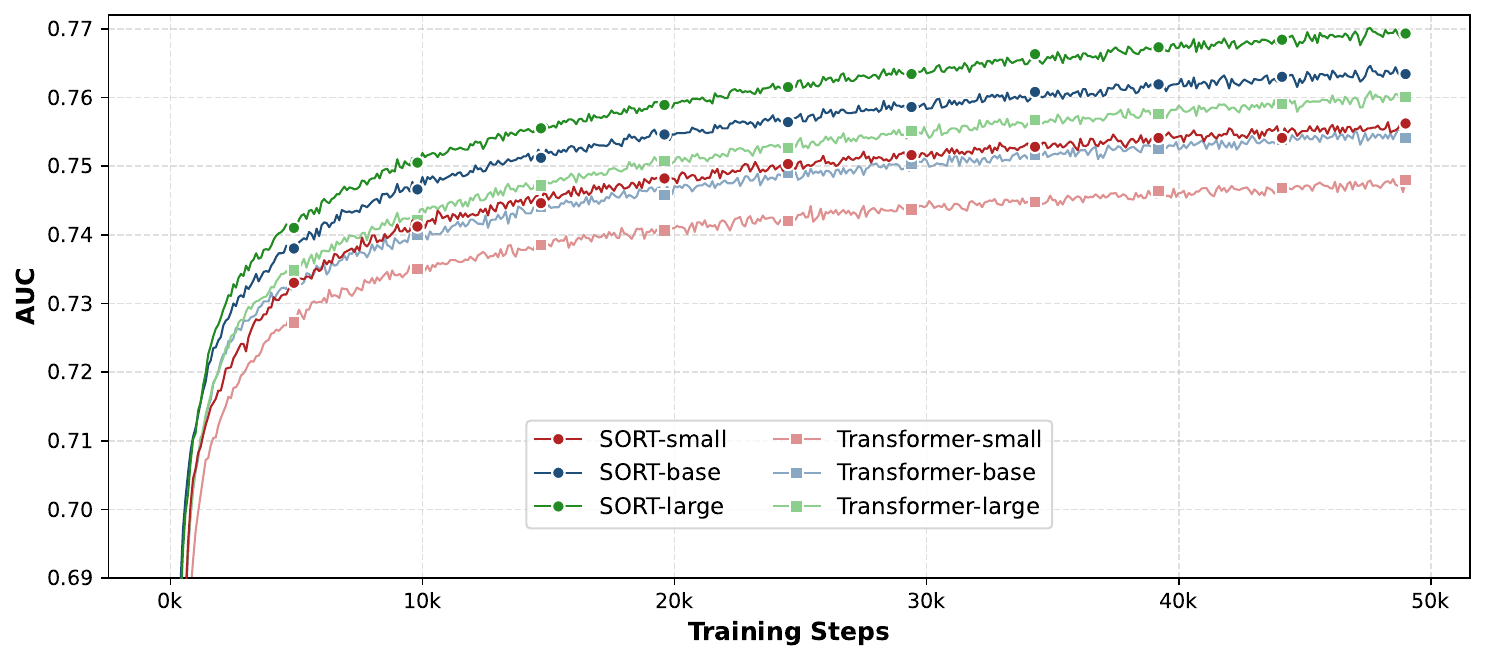}
    \caption{Training AUC curves for CTR task.}
    \label{fig:ctr_auc_curves}
\end{figure}

Beyond static metrics, the robust scalability of SORT is a highlight of our empirical findings. Across small, base, and large scales, SORT consistently maintains a significant advantage over the Transformer baseline. 
Notably, at the large scale, SORT achieves +0.51pt improvement in CTR-AUC, while requiring substantially fewer FLOPs (188G vs. 322G).
As illustrated in Figure~\ref{fig:ctr_auc_curves}, across all scales, the training AUC curves for SORT consistently stay above their Transformer counterparts, demonstrating both its remarkable performance and sample efficiency.

\begin{table*}
    \scalebox{0.85}{
    \begin{tabular}{llcccccccc}
        \toprule
        \multirow{2}{*}{Scale}&\multirow{2}{*}{Model}&\multicolumn{2}{c}{CTR}&\multicolumn{2}{c}{CVR}&\multicolumn{2}{c}{AddCart}& Params&FLOPs \\ 
        \cline{3-8}
        &&AUC&Imp&AUC&Imp&AUC&Imp \\ 
        \midrule
        \multirow{10}{*}{Base}&\makecell[l]{Transformer}&0.7456&-&0.9093&-&0.8663&-&18M&43G \\
        % \hline
        &\makecell[l]{HSTU}&0.7438&-0.18pt&0.9070&-0.23pt&0.8644&-0.19pt&18M&43G \\
        &\makecell[l]{OneTrans}&0.7476&+0.20pt&0.9103&+0.10pt&0.8677&+0.14pt&54M&24G \\
        \cline{2-10}
        &\makecell[l]{Transformer+special token (ST)}&0.7489&+0.33pt&0.9113&+0.20pt&0.8685&+0.22pt&18M&43G \\
        &\makecell[l]{Transformer+local attention (LA)}&0.7466&+0.10pt&0.9099&+0.06pt&0.8671&+0.08pt&18M&40G \\
        &\makecell[l]{Transformer+query pruning (QP)}&0.7482&+0.26pt&0.9112&+0.19pt&0.8676&+0.13pt&18M&24G \\
        &\makecell[l]{Transformer+attention gate (AG)}&0.7477&+0.21pt&0.9097&+0.04pt&0.8672&+0.09pt&20M&47G \\
        &\makecell[l]{Transformer+QKNorm (QKN)}&0.7468&+0.12pt&0.9104&+0.11pt&0.8670&+0.07pt&18M&43G \\
        &\makecell[l]{Transformer+MoE}&0.7475&+0.19pt&0.9099&+0.06pt&0.8675&+0.12pt&83M&43G \\
        \cline{2-10}
        &\makecell[l]{SORT(Transformer+ST+LA+QP+AG+QKN+MoE)}&\textbf{0.7497}&\textbf{+0.41pt}&\textbf{0.9118}&\textbf{+0.25pt}&\textbf{0.8694}&\textbf{+0.31pt}&85M&24G \\
        \bottomrule
        
        \multirow{2}{*}{Small}&\makecell[l]{Transformer}&0.7424&-&0.9049&-&0.8612&-&3M&8G \\
        \cline{2-10}
        &\makecell[l]{SORT}&\textbf{0.7465}&\textbf{+0.41pt}&\textbf{0.9082}&\textbf{+0.33pt}&\textbf{0.8656}&\textbf{+0.44pt}&14M&4G \\
        \bottomrule
        
        \multirow{2}{*}{Large}&\makecell[l]{Transformer}&0.7485&-&0.9122&-&0.8695&-&144M&322G \\
        \cline{2-10}
        &\makecell[l]{SORT}&\textbf{0.7536}&\textbf{+0.51pt}&\textbf{0.9156}&\textbf{+0.34pt}&\textbf{0.8736}&\textbf{+0.41pt}&685M&188G \\
        \bottomrule
    \end{tabular}
    }
    \caption{Comparison of different models. Results are grouped by model scale for fair comparison. Bold indicates the best performance within each group. Imp stands for absolute improvements over the Transformer baseline within each scale. Note that FLOPs are calculated for a single sample forward pass, and Params refers to the number of parameters excluding embeddings. For all models, the number of embedding parameters is about 20B.}
    \label{tab:main}
\end{table*}

\subsection{Data, Model and Sequence Length Scaling}
\label{sec:scaling}

We conducted scaling experiments across three dimensions: data size, model size (number of parameters), and sequence length. 
Unlike LLMs, which benefit from a virtually inexhaustible supply of data, recommender systems are inherently constrained by the volume of user interactions. Consequently, the available data within a given timeframe is finite. To address this limitation, we explore the following two strategies to expand the effective data size.
The first method is to train the model for multiple epochs on the same dataset. Benefiting from generative pre-training and frozen embeddings, the model avoids the common issue of overfitting during multiple epochs, as we see in Figure~\ref{fig:multi_epoch_training}. The second approach involves combining data from other scenarios with the target set to improve model performance on the target scenario. In this setup, the specific scenario identifier is encoded as an input feature, enabling the model to distinguish between different data sources.
In summary, we investigate the scaling effects under the following experimental configurations:
\begin{itemize}
    \item Data scaling (multi-epoch): Training for 1, 2, and 3 epochs.
    \item Data scaling (multi-scenario): Comparison between single-scenario and multi-scenario settings.
    \item Model scaling: Model variants of small, base, and large sizes.
    \item Sequence length scaling: Sequence lengths of 256, 512, 1K, 2K and 4K.
\end{itemize}

\begin{figure*}
    \centering
    \includegraphics[width=0.85\textwidth]{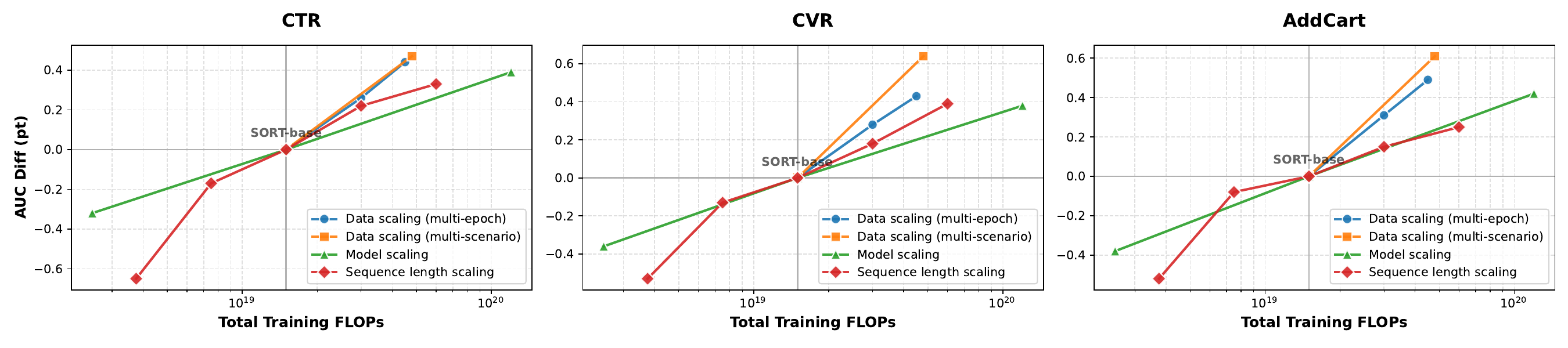}
    \caption{Scaling SORT over data size, model size and sequence length. The SORT-base model is trained on single-scenario data for one epoch, with sequence length of 1K.}
    \label{fig:scaling}
\end{figure*}

The experimental results are illustrated in Figure~\ref{fig:scaling}. We observe that increasing the data size, model size, or sequence length consistently leads to performance gains, thereby validating the strong scalability of our proposed method. Overall, data scaling proves to be the most effective, exhibiting a steeper upward slope in performance improvement.
This finding suggests that in production, the best practice is to assign higher priority to data scaling, followed by the sequence length scaling and model scaling, to achieve better performance under a given computational budget.

\subsection{Online A/B Testing}

\begin{table}
    \caption{
    Online A/B test results over a one-month duration. 
    % All metrics represent relative improvements over the production baseline.
    % Online A/B test results on multiple recommendation scenarios.
    $\uparrow$ ($\downarrow$) indicates higher (lower) values are preferred.}
    \scalebox{0.80}{
        \begin{tabular}{lcccccc}
            \toprule
            Scenario & Orders$\uparrow$ & Buyers$\uparrow$ & GMV$\uparrow$ & Latency$\downarrow$ & Throughput$\uparrow$ \\
            \midrule
            HomePage & +2.37\% & +2.58\% & +10.01\% & -62\% & +325\% &  \\
            DetailPage & +8.99\% & +7.46\% & +11.49\% & -51\% & +405\% & \\
            ShoppingCart & +7.10\% & +6.74\% & +6.98\% & -64\% & +448\% & \\
            AfterPayment & +11.43\% & +9.91\% & +6.12\% & -72\% & +1180\% & \\
            \midrule
            Macro Average & +7.47\% & +6.67\% & +8.65\% & -62\% & +589\% & \\
            \bottomrule
        \end{tabular}
    }
    \label{tab:online_result}
\end{table}

To validate the effectiveness of SORT, we conduct extensive online A/B testing in four main recommendation scenarios on AliExpress, including HomePage, DetailPage, ShoppingCart and AfterPayment.
The production baseline utilizes the conventional DLRM architecture, which has been incrementally trained on data spanning over a year.
In contrast, the deployed SORT model, which employs a base-scale configuration, is trained using only recent three months' data.

The experimental results (shown in Table~\ref{tab:online_result}) demonstrate that SORT achieves significant improvements in both effectiveness and efficiency.
In terms of effectiveness, SORT yields substantial gains across multiple key performance indicators across all scenarios. On average, SORT achieves remarkable improvements of 7.47\%, 6.67\%, and 8.65\% in orders, buyers, and GMV, respectively.
This highlights the ability of SORT to accurately model user interests and improve the core metrics of the recommender system.
Regarding serving efficiency, SORT achieves an average 62\% reduction in latency and an average 589\% boost in throughput in four scenarios.
The improved efficiency is attributed to a synergy of data, model, and system-level optimizations. Specifically, the request-centric data paradigm enables the reuse of shared user profile and historical sequence across all candidates in a single pass. Furthermore, the unified Transformer architecture mitigates the computational overhead associated with fragmented operators. These factors, combined with our optimized inference system, collectively result in significant gains in efficiency. 
% The SORT model has been rolled out to all users on the platform.

\section{Conclusion}
We present SORT, a systematically optimized ranking Transformer that bridges the gap between Transformers and industrial-scale ranking. Targeting high feature sparsity and low label density, SORT combines request-centric sample organization, local attention with query pruning, generative pre-training, and architectural refinements (special tokens, QKNorm, attention gate, sparse MoE) into a unified, hardware-efficient design. SORT outperforms strong baselines and scales smoothly along data, model size and sequence length, with data scaling yielding the steepest gains under a fixed compute budget. Fully deployed on AliExpress, SORT delivers substantial business gains (+8.65\% GMV) while cutting latency by 62\% and boosting throughput nearly sevenfold.

\section{GenAI Usage Disclosure}
Generative AI tools (Qwen, Gemini) were used solely for grammatical editing/polishing of the text. All content, ideas, and data analysis remain the original work of the authors.

% \begin{acks}
% \end{acks}

%%
%% The next two lines define the bibliography style to be used, and
%% the bibliography file.
\bibliographystyle{ACM-Reference-Format}
\balance
\bibliography{ranknext}
% \onecolumn % 切换为单栏
\appendix

\end{document}